\documentstyle[preprint,aps]{revtex}

\newcommand{\bm}[1]{\mbox{\boldmath $ #1  $}}
\newcommand{\kg}{\; ^< \hspace{-.22cm} _\sim \;}

\newcommand{\be}{\begin{equation}}
\newcommand{\ee}{\end{equation}}
\newcommand{\mb}[1]{\mbox{\scriptsize #1}}

\newcommand{\hs}[1]{\hspace{- \arraycolsep}}

\newcommand{\mx}{\mbox}

\newcommand{\bi}{\bigskip}
\newcommand{\no}{\noindent}

\begin{document}

\draft

\title{Relativistic ponderomotive force, uphill acceleration, 
and transition to chaos}

\author{D. Bauer, P. Mulser, and W.-H. Steeb$^*$}

\address{Theoretical Quantum Electronics (TQE), Technische Hochschule Darmstadt, 
Hochschulstr. 4A, D-64289 Darmstadt, FRG}

\date{{\sc Reprint of} Phys.~Rev.~Lett. {\bf 75} 4622 (1995)}

\maketitle

\begin{abstract}
Starting from a covariant cycle-averaged Lagrangian the relativistic 
oscillation center equation
of motion of a point charge is deduced and 
analytical formulae for the ponderomotive force in a travelling wave of 
arbitrary strength are presented. It is further shown that the ponderomotive 
forces for transverse and longitudinal waves are different; in the latter, 
uphill acceleration can occur. In a standing wave there exists a threshold
intensity above which, owing to transition to chaos, the secular motion can 
no longer be described by a regular ponderomotive force. 
\end{abstract}

\pacs{PACS number(s): 52.20.Dq,05.45.+b,52.35.Mw,52.60.+h}

\narrowtext

A so-called ponderomotive potential
$\Phi_p$ is induced by the oscillatory motion of a free charge. This
potential plays a dominant role in atomic physics (e.g.\ 
multiphoton ionization) and laser plasma dynamics (e.g.\ parametric 
instabilities, self focusing, beat wave accelerator, fast ignitor
\cite{eins}).
It was shown independently by several authors \cite{zwei} that for a monochromatic
electromagnetic field of arbitrary space dependence,
\be
{\bm E} ({\bm x},t) = \Re \hat{\bm E} ({\bm x}) e^{- i \omega t},
\ee
the oscillation center
dynamics of a charge $q$ is governed by the so-called ponderomotive force ${\bm f}_p $,
\be
{\bm f}_p = - \frac{q^2}{4 m \omega^2} {\bm \nabla} ( \hat{\bm E}\cdot \hat{\bm E}^*) .
\ee
Eq.(2) was obtained from a first order perturbation analysis of the 
Lorentz force \cite{drei} around the oscillation center and is therefore subject to 
the usual smallness constraints of certain parameters. In order to obtain some
weak generalizations of this expression, a variety of different approaches was
chosen [4--6]. Again, they
are characterized by (i) a 
perturbation analysis of momentum, (ii) harmonic fields of type Eq.(1), and 
(iii) a small ratio of oscillation amplitude to wavelength $ \lambda$. 
Recently, however, in connection with the existence of new lasers capabable 
of delivering ultrahigh irradiance a real need for relativistic expressions for 
${\bm f}_p $, not bound by such limits, has arisen. One of the aims of this 
letter is to show under which conditions ${\bm f}_p$ exists and what forms 
it assumes.


The relativistic Lagrangian $ L ({\bm x}, {\bm v}, t), \;\; {\bm v} = d 
{\bm x}/dt$, of a charge $q$ in an arbitrary electromagnetic field $ 
{\bm E} = - {\bm \nabla} \Phi - \partial {\bm A}/\partial t $ is given by 
\be
L({\bm x}, {\bm v}, t) = - \frac{m c^2}{\gamma} + q {\bm v}\cdot {\bm A} - q \Phi;
\;\; \gamma = ( 1 - v^2/c^2)^{- 1/2} .
\ee
When an oscillation center exists, the transformation to action-angle 
variables $ S = S ({\bm x}, t), \;\; \eta = \eta ({\bm x}, t) $ is
possible
(e.g.\ $\eta={\bm{k}\cdot \bm{x}}-\omega t$ in the case of a travelling
monochromatic wave). 
The 
action $S$ and the angle $\eta $ are both Lorentz-invariant. The motion of 
the particle is governed by Hamilton's principle,
\be
\delta S = 
\delta 
\int^{\eta_2}_{\eta_1} L ({\bm x} (\eta), {\bm v} (\eta), t (\eta)) \frac{dt}{
d \eta} d \eta = 0.
\ee
From the Lorentz invariance of $S$ and $\eta$ follows that the Lagrangian 
${\cal L} (\eta) = L (d \eta / dt)^{-1} $ is invariant with respect to a 
change of the inertial reference system. Assuming that $\eta$ is normalized to 
$ 2 \pi $ for one full cycle or period of motion, the cycle-averaged 
Lagrangean ${\cal L}_0 $,
\be
{\cal L}_0 (\eta) = \frac{1}{2 \pi} \int^{\eta + 2 \pi}_{\eta} {\cal L}
(\eta') d \eta',
\ee
depending only on the secular (i.e.\ oscillation center) coordinates 
${\bm x}_0, {\bm v}_0 $ through $\eta$ is defined. The oscillation center 
motion is governed by the Lagrange equations of motion, 
\be
\frac{d}{dt} \frac{\partial L_0}{\partial {\bm v}_0} - 
\frac{\partial L_0}{\partial {\bm x}_0} = 0,
\ee
with $L_0={\cal L}_0 d\eta/dt$.
To demonstrate this assertion we prove the following \\
{\bf Theorem:} The validity of Eq.(4) implies
\be
\delta \int^{\eta_f}_{\eta_i} {\cal L}_0 (\eta) d \eta = o (N^{-1} ).
\ee
Thereby $ N = (\eta_f - \eta_i)/ 2 \pi $ is the number of cycles over which
${\cal L}_0 $ undergoes an essential change. The 
symbol $o(N^{-1}) $ means: ''vanishes at least with order $ 1 / N$``. \\
{\bf Proof}: Let the variation be an arbitrary piecewise continuous function
$\Delta (\eta) $. The $n$-th cycle starts at $ \eta = \eta_n$, at which for
brevity we use the symbols $ \Delta_n = \Delta (\eta_n), \linebreak \partial 
{\cal L}_0/\partial \eta_n = (\partial {\cal L}_0 / \partial \eta)_{\eta =
\eta_n}$. If the same quantities refer to an intermediate point $ \eta \leq 
\eta_a \leq \eta_n + 2 \pi $ we write $ \Delta_a $ and $ \partial {\cal L}_0
/ \partial \eta_a$ and omit the index $n$ for the interval. In leading
order holds:
\begin{eqnarray*}
\lefteqn{\left| \delta \int^{\eta_f}_{\eta_i} {\cal L}_0 d \eta
\right| 
= \left| \int^{\eta_f}_{\eta_i} \delta 
({\cal L}_0 - {\cal L}) d \eta \right|}  
\\
&=& \left| \int^{\eta_f}_{\eta_i} \left[ {\cal L}_0 (\eta + \Delta) - {\cal L}
(\eta + \Delta) \right] d \eta - \int^{\eta_f}_{\eta_i} \left[ {\cal L}_0 
(\eta) - {\cal L} (\eta) \right] d \eta \right|
\\
&\leq& \sum_n \left| \int^{\eta_n + 2 \pi}_{\eta_n} {\cal L}_0 (\eta + \Delta)
d \eta - 2 \pi {\cal L}_0 ( \eta_n + \Delta_n) - \left[  \int^{\eta_n + 2 
\pi}_{\eta_n} {\cal L}_0 (\eta) d \eta - 2 \pi {\cal L}_0 (\eta_n ) \right]
\right|
\\
&=& \sum_n \left| \int^{\eta_n + 2 \pi}_{\eta_n} \frac{\partial {\cal L}_0}{
\partial \eta_n} \Delta (\eta) d \eta - 2 \pi \frac{\partial {\cal L}_0}{
\partial \eta_n} \Delta_n \right| = 2 \pi \sum_n \left| \frac{\partial {\cal 
L}_0}{\partial \eta_a} \Delta_a - \frac{\partial {\cal L}_0}{\partial \eta_n} 
\Delta_n \right|.
\end{eqnarray*}
In the last step the mean value theorem is used. The function $ \Delta (\eta)
$ is arbitrary. Therefore at $ \eta = \eta_n \;\; \Delta_n = \Delta_a $ can
be chosen now without affecting $ \Delta_a$. With this substitution
the leading order gives the result
\[
\left| \delta \int^{\eta_f}_{\eta_i} {\cal L}_0 d \eta \right| \leq (2 \pi)^2
\sum_n \left| \frac{\partial^2 {\cal L}_0}{\partial \eta^2_n} \Delta_a \right|
\leq (2 \pi)^2 N \max \left| \frac{\partial^2 {\cal L}_0}{\partial \eta^2_n} 
\right| \times \max | \Delta_a |.
\]
In this last step it is essential that $ \partial {\cal L}_0 / \partial \eta $ 
is a smooth function (in contrast to $ \Delta $ which is generally not). Now,
$ N = \min \frac{1}{2 \pi} |{\cal L}_{0\max}/(\partial {\cal L}_0/\partial \eta_n)
| $ is chosen, i.e.\ over $N$ cycles $ {\cal L}_{0 \max} $ changes at most by ${\cal 
L}_0$. It follows that 
\be
\left| \delta \int^{\eta_f}_{\eta_i} {\cal L}_0 d \eta \right| \leq 
\frac{ |{\cal L}_{0 \max} |}{N} \times \max | \Delta |.
\ee
Performing the variation of this inequality leads to Eq.(6) with the $0$ 
replaced by a function $f$ not larger than $ \left| {\cal L}_{0 \max} \right|
\times \max |\Delta| / N^2$ . \hfill $\bullet$ 

In order to understand what inequality (8) means let
us specialize to a case of the averaged Lagrangian ${\cal L}_0$ not 
depending explicitly on time. Then the Hamiltonian $ H_0 = {\bm p}_0\cdot {\bm v}_0
- L_0 ({\bm x}_0, {\bm v}_0)$, where $L_0 = - {\cal L}_0 (\omega -
{\bm k} \cdot 
{\bm v}_0)$, owing to $ dH/dt = \partial H / \partial t = -  
\partial L_0 / \partial t = 0 $, expresses the energy conservation $ H = E 
= \mx{const}$. A straightforward estimate shows that the uncertainty $f$ in 
Eq.(6) leads to an energy uncertainty $ \Delta H / H \kg 2 \pi/N$. This 
means that Eq.(6) is adiabatically zero and the total cycle-averaged energy 
is an adiabatic invariant in the rigorous mathematical sense \cite{sieben}. For $N 
\rightarrow \infty $ Eq.(6) becomes exact. Perturbative averaging of a
Lagrangian was used in \cite{kentw}. 


The relativistic Hamiltonian of a point charge in the electromagnetic field 
follows from Eq.(3),
\be
H = {\bm p}\cdot {\bm v} - L = \left\{ m^2 c^4 + c^2 ( {\bm p} - q {\bm A})^2 
\right\}^{1/2} + \Phi,
\ee
with the canonical momentum $ {\bm p} = \partial L/\partial {\bm v} = \gamma 
m {\bm v} + q {\bm A}$. Its numerical value is the total energy $ E = \gamma 
m c^2 + \Phi $. Considering a monochromatic wave in vacuum we can set
$ \Phi = 0 $.  This motion is exactly solvable \cite{acht}. The cycle-averaged quiver energy in the 
oscillation center system is given by 
\be
H_0 - m c^2 
= mc^2 \left\{
\left( 1 + \frac{q^2}{\alpha m^2c^2} \hat{\bm A}\cdot \hat{\bm A}^*\right)^{1/2}-
1 \right\} = W
\ee
with $ \alpha = 1 $ for circular and $ \alpha = 2 $ for linear polarization. 
If the effective mass $ m_{\mb{eff}} = -{\cal L}_0 \gamma_0
(d\eta/dt)/c^2$ is introduced, $L_0={\cal L}_0 d\eta/dt$ shows that in
an arbitrary inertial frame in which the oscillation center  moves at
speed $ {\bm v}_0, \;\; L_0$ and $ H_0 $ are those of a free particle
with space and time dependent mass $m_{\mb{eff}} $,  
\be
L_0 ({\bm x}_0, {\bm v}_0, t) = - \frac{m_{\mb{eff}} c^2}{\gamma_0}, \;\; H_0 
({\bm x}_0, {\bm p}_0, t) = \gamma_0 m_{\mb{eff}} c^2;
\ee
\[
\gamma_0 = \left( 1 - \frac{v_0^2}{c^2} \right)^{-1/2}, \;\; p_0 = \gamma_0
m_{\mb{eff}} {\bm v}_0.
\]
Expressions (11) hold in any electromagnetic field in vacuum in which an 
oscillation center can be defined. In the special case of Eq.~(10)
$m_{\mb{eff}} = (1+q^2\hat{\bm A}\cdot \hat{\bm A}^*/\alpha m^2c^2)^{1/2}$.

In the oscillation center system, i.e.\ in the inertial frame in which at the 
instant $t \;\; {\bm v}_0 (t) = 0 $ holds, the ponderomotive force follows
from Eqs.(6) and (11),
\be
{\bm f}^N_p \equiv \frac{d {\bm p}_0}{d \tau} =  \frac{\partial L_0}{\partial
{\bm x}_0} = - c^2 {\bm \nabla} m_{\mb{eff}}.
\ee
The force ${\bm f}_p $ is given the index $N$ since it is a Newton force 
(co-moving system); $\tau$ is the proper time. The Minkowski ponderomotive force $ F_p$, valid in any inertial 
system, is
\be
F_p 
= 
\left( {\bm f}^N_p + \frac{\gamma_0 -1}{v_0^2} ({\bm f}^N_p\cdot {\bm v}_0) 
{\bm v}_0, \;\; \frac{\gamma_0}{c} {\bm v}_0\cdot {\bm f}^N_p \right).
\ee
Hence, the three-dimensional relativistic ponderomotive force ${\bm f}_p $ at 
any oscillation center speed $ {\bm v}_0 $ is the Einstein force,
\be
{\bm f}_p 
= - \frac{c^2}{\gamma_0} \left\{ {\bm \nabla}^N m_{\mb{eff}} + \frac{\gamma_0 -1}{
v_0^2} ({\bm v}_0\cdot {\bm \nabla}^N m_{\mb{eff}}) {\bm v}_0 \right\}.
\ee

\no In the non-relativistic limit Eq.(2) is easily recovered. 

To see the power of the Lagrangian formulation, Eq.(6), we calculate 
${\bm f}_p $ in a non-relativistic Langmuir wave of the form $E (x,t) =
\hat{E} (x,t) \sin (kx - \omega t) $ with slowly varying amplitude 
$\hat{E} $. In lowest order the potential is $ \Phi (x,t) =  
(\hat{E}/k) \cos (kx - \omega t)$ and $ L = m v^2/2 - q \Phi$. In the 
frame comoving with $ x_0 $ the particle sees the Doppler-shifted frequency
$ \Omega = \omega - k v_0$ ( plus higher harmonics which are not essential here).
With the periodic excursion $ \zeta (t) $ around $ x_0 $ the potential is 
$ \Phi (x,t) \simeq \Phi (x_0, t) + \zeta (t) \partial \Phi/\partial x_0$. 
From this $ L_0 = m v_0^2/2 - \alpha \hat{E}^2/ \Omega^2, \;\; \alpha =
q^2/4m$, results in lowest order. With this Lagrangian it follows from
Eq.(6) for $ \hat{E} = \hat{E} (x)$ (no explicit time dependence) that
\be
f_p = m \frac{dv_0}{dt} 
= -
\alpha \frac{(1 - V_0)(1 - 3 V_0)}{\omega^2 (1-V_0)^4 - 6 \alpha 
\hat{E}^2/m v^2_{\varphi}} \frac{\partial}{\partial x} \hat{E}^2.
\ee

In the last expression $ V_0 $ is the oscillation center velocity 
normalized to the phase velocity $ v_{\varphi} = \omega/k, \;\; V_0 = v_0 /
v_{\varphi}$. The formula shows that $ f_p $ changes sign when the particle 
is injected into the Langmuir wave with a velocity $ v_0 $ exceeding $ 
v_{\varphi}/3 $. In regions where the standard expression for $ f_p$, Eq.(2),
always exhibits repulsion the more exact treatment can lead to
attraction. 
Eq.~(15) was also
derived in a more formal but physically less transparent manner in
\cite{sechs}.  Uphill acceleration in an electron plasma wave of 
increasing amplitude $ \hat{E} (x) $ is confirmed by the numerical 
solution of the exact equation of motion (Fig.1a). If $ \hat{E} (x) $
grows indefinitely the exact ponderomotive force changes sign again and the 
particle stops and is finally reflected; the corresponding path in phase space exhibits a hysteresis (Fig.1b). Uphill acceleration, 
occasionally observed in other context \cite{zehn}, is a general ponderomotive phenomenon.

Strong gradients of amplitude and, as a consequence, large ponderomotive forces
appear when two or more waves superpose. Of particular importance is the case
of a standing, or partially standing wave. We have numerically studied the secular motion in the field given by the
vector potential
\be
{\bm A} (x,t) = \hat{A} {\bm e}_y (\sin \eta_+ - \sin \eta_-), \;\;
\eta _{\pm} = kx \pm \omega t.
\ee


\no
For field amplitudes $ \xi \leq 0.1$ (non-relativistic case), $\xi = |q| \hat{A} / mc$, 
splitting of the motion into a fast and a secular, i.e.\ ponderomotive one, 
makes sense. In Fig. 2a an electron orbit is followed over 110 laser cycles in 
the standing wave of strength $ \xi = 0.1: $ the motion is regular and its 
dependence on the phase of the wave field is weak (narrow band of orbits). 
The corresponding Nd laser intensity $(\omega_{\mb{Nd}} = 1.78 \times 10^{15} 
\; \mx{s}^{-2})$ is $ I_{\mb{Nd}} = 1.21 \times 10^{16} \; \mx{Wcm}^{-2} $.
At $ \xi = 0.25 \; ( I_{\mb{Nd}} = 7.6 \times 10^{16} \; \mx{Wcm}^{-2})$ some
electrons starting near the maximum $ A(x) = 2 \hat{A}$ are able to 
escape from the ponderomotive potential well; the secular motion through the
well exhibits a strong phase dependence. At $ \xi = 0.5 \;\;(I_{\mb{Nd}} = 3 
\times 10^{17} \; \mx{Wcm}^{-2})$ the motion becomes totally chaotic (see Fig.
2b): slight changes in the initial time produce totally different orbits, 
a clear signature of chaos.  The chaotic motion has its origin in the Doppler effect since the
moving electron ''sees`` one wave as blue- and the other one as red-shifted. No 
oscillation center exists under such circumstances and a secular time
scale may only build up on the statistical average.

Electrons in a dense plasma behave differently since they are coupled
to the massive ions by an ambipolar electric field $E_s$. 
In Fig. 3 the calculation of Fig. 2 is repeated for 
$ \xi = 0.5 $ in a plasma of which the density is assumed to be such as to 
keep the oscillation center in a steady state position. In Fig. 3a the single
arc-like (no longer 8-shaped) orbits are shown. The inclusion of $
E_s$ in the equation of motion is essential for producing the regular orbits
and the correct ponderomotive forces (Fig. 3b). For 
comparison the bare dotted line is the ponderomotive force from Eq.(2) 
without including the space charge field $E_s$ in $f_p$. We estimate from numerical 
runs that beyond $ I_{\mb{Nd}} = 10^{18} \; \mx{Wcm}^{-2}$ no regular 
ponderomotive force exists in the dense plasma either. 

We conclude as follows. (i) When an oscillation center of motion exists the
invariant cycle-averaged Lagrangian describes the ponderomotive motion in arbitrarily 
strong fields. (ii) The ponderomotive force in a monochromatic travelling, 
locally plane wave of arbitrary strength in any reference system can be 
expressed analytically. (iii) In a longitudinal wave
uphill acceleration
and phase space-hysteresis may occur. Finally, (iv) superposition of 
different modes of the same frequency leads to a limiting intensity above 
which transition to chaos occurs. 
The influence of dissipation (radiation
losses, friction) on the ponderomotive force is also well understood now, but
will be treated in a separate paper.

\bi
$^*$) On leave from Rand Afrikaans University, Johannesburg, South Africa.

\begin{figure}[hbt]


 \caption{ Ponderomotive motion of an electron ($q=-e$) in a travelling Langmuir 
wave. In (a) the velocity $ V = v/v_{\varphi}, \;\; v_{\varphi} = \omega/k $,
is shown as a function of position $ X = kx$ in the longitudinal field
$ \xi (1-\exp(-0.1 X)) \sin (X-T), \;\;  \xi = e \hat{E}/m \omega v_{\varphi}
= 0.1$, for the initial conditions
$X_0 = 0, \;\; V_0 = 0.6 $ (trajectory a). The oscillation center motion from 
Eq. (2) (dotted line b) shows a slight decrease, whereas the secular 
acceleration by Eq. (15) (dotted line c) is in good agreement with that 
resulting from trajectory a. The lower curve d shows the harmonic electric 
field as ''seen`` by the moving electron. In (b) the normalized kinetic 
energy $V_0^2/2$ is shown as a function of $ \Phi_p/m_e v_{\varphi}^2 $ from Eq. 
(2) for the linearly 
increasing field $\; 2.5 \times 10^{-3} \xi X \sin (X-T), \;\; \xi = 1$, and 
injection velocities $ V_0 = 1/3 $ (lower curves) and $ V_0 = 0.43 $ (upper
curves). The motion shows a pronounced hysteresis in phase space owing to 
different acceleration in co- and counter-motion. 
}

\end{figure}

\begin{figure}[hbt]


 \caption{ Regular (a) and stoachstic motion (b) over 110 laser cycles of 
an electron placed with oscillation center speed $ v_0 = 0 $ at the position
$ X = kx_0 = 39 \pi/40 $ in a plane monochromatic standing wave; ordinate $ 
Y = ky $ points into field direction. In (a) the maxima of the ponderomotive
potential $ \Phi_p $ are located at $ X = 0, \;\; \pi$; at $ X = \pi/2 \;\;
\Phi_p = 0$; normalized amplitudes of counterpropagating waves are $ \xi = e
\hat{A}/mc = 0.1$. The electron motion evolves in a narrow band between the 
potential maxima. Band broadning around $ X = \pi/2 $ is due to Doppler 
effect. In (b) $\Phi_p$ (dashed line) is 25 times stronger $(\xi = 0.5)$, 
the electron motion is erratic, sweeping over 21 maxima of $\Phi_p $ during 
110 cycles. Note also the enormous transverse excursion of $ \Delta y = 0.9 $ 
wavelengths $(\Delta Y = 5.8)$.
}

\end{figure}

\begin{figure}[hbt]


 \caption{ Electron orbits and ponderomotive potential in a standing plane,
monochromatic wave in a uniform plasma. (a) Electron orbits around oscillation
center tied to the ions by the induced ambipolar field; field distribution
as in Fig. 2(a), field strength $\xi = 0.5 $ as in Fig. 2(b). In (b) the 
segments parallel to $ X = kx $ indicate the orbit widths in $X$-direction;
the connecting dotted line indicates the relativistic ponderomotive force.
For comparison, the bare dotted line is $f_p$ from Eq. (2); the field is 
given in dimensionless units of $ eE/m \omega c$.
}

\end{figure}





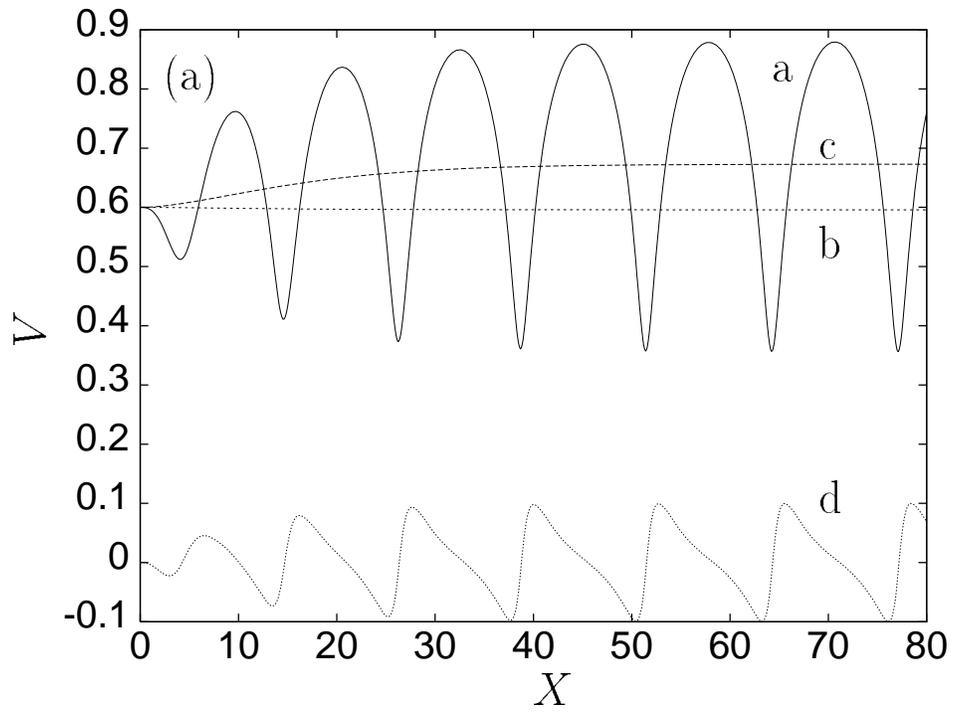
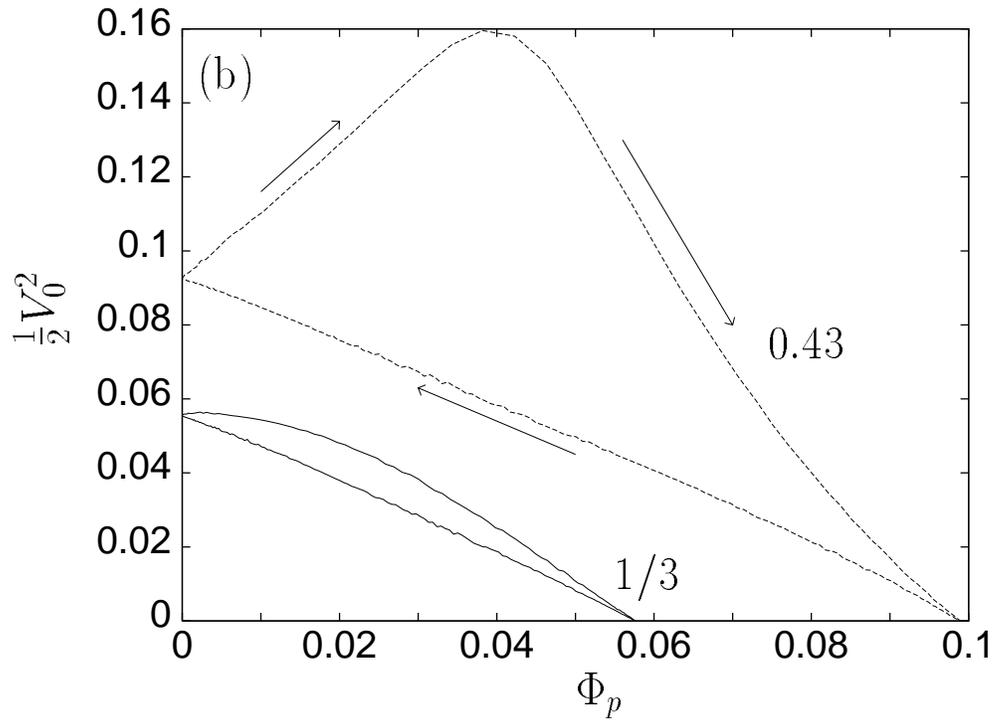
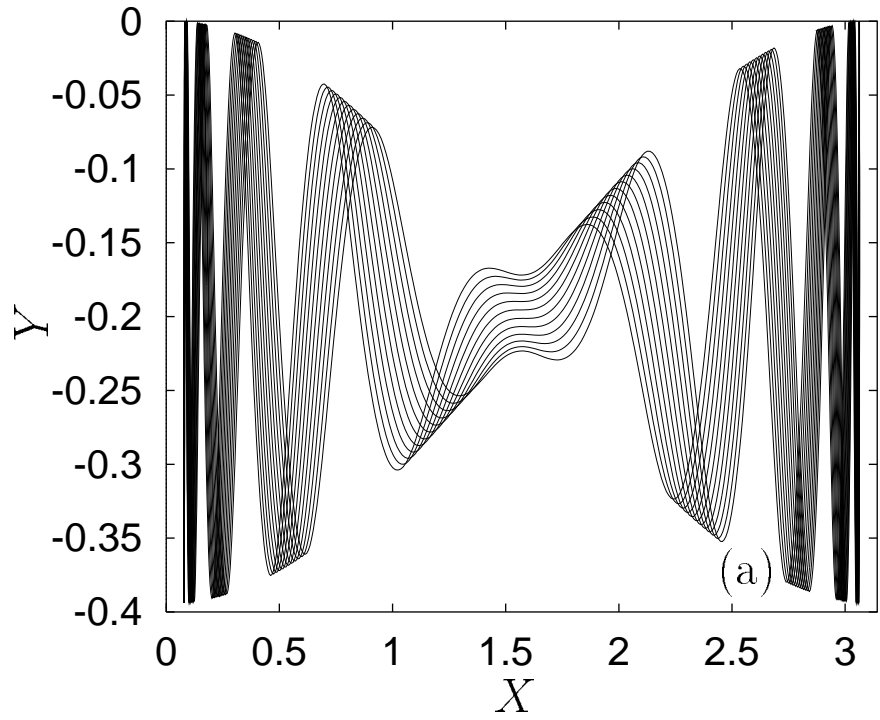
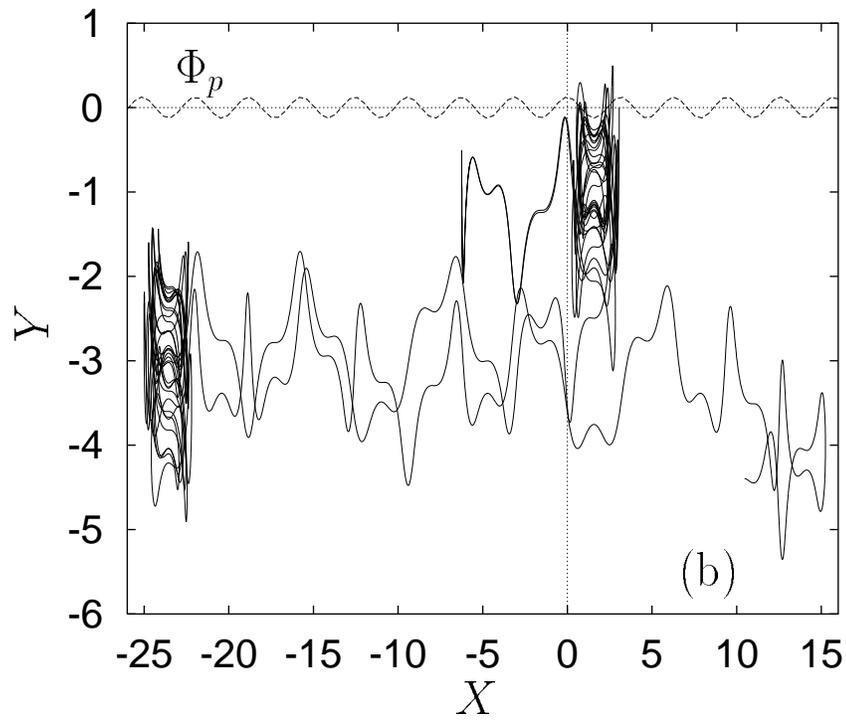
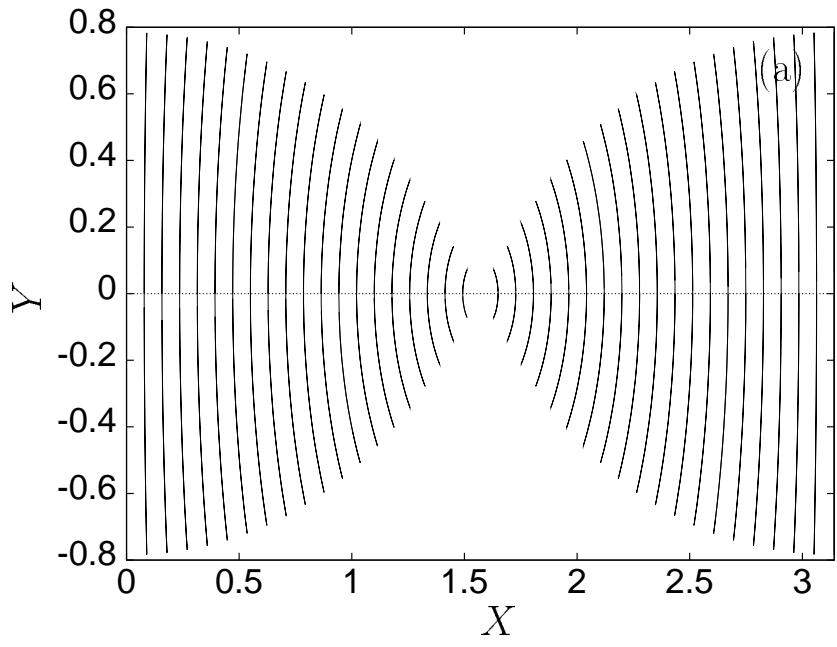
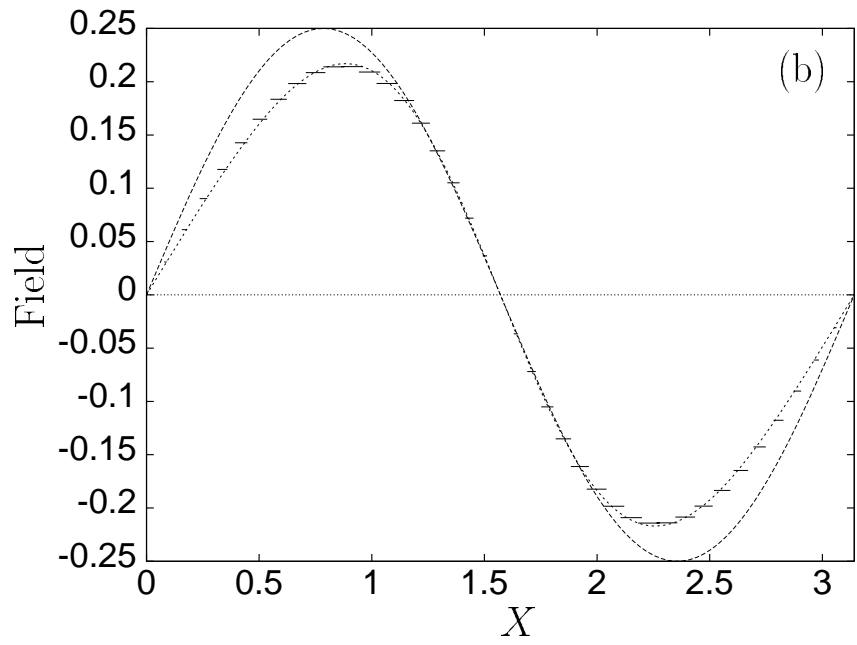
\end{document}